\documentclass[usegraphicx]{mn2e}

\usepackage{amssymb}
\usepackage{mathptmx}
\usepackage{url}

\newcommand{\Mpc}{\rm\thinspace Mpc}
\newcommand{\kpc}{\rm\thinspace kpc}

\newcommand{\km}{\rm\thinspace km}

\newcommand{\cm}{\rm\thinspace cm}

\newcommand{\pcmcu}{\hbox{$\cm^{-3}\,$}}

\newcommand{\yr}{\rm\thinspace yr}

\newcommand{\s}{\rm\thinspace s}

\newcommand{\Msun}{\hbox{$\rm\thinspace M_{\odot}$}}

\newcommand{\Msunpyr}{\hbox{$\Msun\yr^{-1}\,$}}

\newcommand{\erg}{\rm\thinspace erg}

\newcommand{\ergps}{\hbox{$\erg\s^{-1}\,$}}

\newcommand{\kmps}{\hbox{$\km\s^{-1}\,$}}

\newcommand{\kmpspMpc}{\hbox{$\kmps\Mpc^{-1}$}}

\newcommand{\Zsun}{\hbox{$\thinspace \mathrm{Z}_{\odot}$}}

\newcommand{\pcm}{\hbox{$\cm^{-3}\,$}}
\newcommand{\psqcm}{\hbox{$\cm^{-2}\,$}}

\begin{document}
\title{A \emph{Chandra} observation of the disturbed cluster core of
  Abell 2204}

\author[J.S.  Sanders, A.C. Fabian and G.B. Taylor]{J.S.
  Sanders${}^1$\thanks{E-mail: jss@ast.cam.ac.uk},
  A.C. Fabian${}^1$ and G.B. Taylor${}^2$\\
  ${}^1$ Institute of Astronomy, Madingley Road, Cambridge. CB3 0HA\\
  ${}^2$ National Radio Astronomy Observatory, P.O. Box 0, Socorro, NM
87801, USA}
\maketitle

\begin{abstract}
  We present results from an observation of the luminous cluster of
  galaxies Abell 2204 using the \emph{Chandra X-ray Observatory}.  We
  show the core of the cluster has a complex morphological structure,
  made up of a high density core ($n_e \sim 0.2\pcm$) with flat
  surface brightness, a surrounding central plateau, a tail-like
  feature, wrapping around to the east, and an unusual radio source. A
  temperature map and deprojected-profile shows that the temperature
  rises steeply outside these regions, until around $\sim 100$~kpc
  where it drops, then rises again. Abundance maps and profiles show
  that there is a corresponding increase in abundance at the same
  radius as where the temperature drops. In addition there are two
  cold fronts at radii of $\sim 28$ and 54.5 kpc. The disturbed
  morphology indicates that the cluster core may have undergone a
  merger.  However, despite this disruption the mean radiative cooling
  time in the centre is short ($\sim 230$~Myr) and the morphology is
  regular on large scales.
\end{abstract}


\section{Introduction}
The cluster of galaxies Abell~2204 lies at a redshift of 0.1523. It is
luminous ($L_X = 2 \times 10^{45} h_{50}^{-2} \ergps$ in the 2-10 keV
band; Edge et al 1990). Schuecker et al (2001) and Buote \& Tsai
(1996) found this cluster had a regular morphology based on
\emph{ROSAT} data.

Peres et al (1998) identified this cluster as having the second most
massive cooling flow within the brightest 50 galaxy clusters. In
post-\emph{Chandra}/\emph{XMM-Newton} times this means it has a highly
peaked soft X-ray surface brightness profile. Using a
constant-pressure cooling flow spectral model Allen (2000) found a
best-fitting cooling flow flux value of $2103^{+356}_{-378}\Msunpyr$.
The central galaxy of Abell~2204 is one of the strongest optical line
emitters within a redshift of 0.2, with much massive star formation
(Crawford et al 1999), CO line emission (Edge 2001) and H${}_2$ line
emission (Edge et al 2002), indicating the presence a significant mass
of cold gas near the centre.

Jenner (1974) found that the two optical nuclei of this cluster differ
in velocity by $249\kmps$.  Clowe \& Schneider (2002) detected weak
lensing gravitational shear around Abell~2204 at high significance,
whilst Dahle et al (2002) observed around a dozen red and blue arcs
and arclets surrounding the central cluster galaxy, and found the mass
distribution to be elongated in the east-west direction.

Distances in this paper assume a cosmology with $H_0 = 70 \kmpspMpc$
and $\Omega_\Lambda = 0.7$. With this cosmology, 1~arcsec corresponds
to a distance of 2.6~kpc. The data were processed with version 3.0.2
of \textsc{ciao}, and the spectra were fit with version 11.3 of
\textsc{xspec} (Arnaud 1996). We used the solar abundance ratios of
Anders \& Grevesse (1989). All positions are in J2000 coordinates.

\section{Data preparation}
Abell~2204 was observed by \emph{Chandra} in 2000-07-29. The data were
reprocessed with the latest gain file appropriate for the dataset
(\texttt{acisD2000-01-29gain\_ctiN0001.fits}). Time dependent gain
correction was also applied using the \textsc{corr\_tgain} utility and
the November 2003 versions of the correction (Vikhlinin 2003).  In
addition the \textsc{ciao} version of the \textsc{lc\_clean} script
(Markevitch 2004) was used to remove periods of the dataset where the
count rate was not quiescent. Very little time was removed, producing
an event file with an exposure of 10.1~ks.  We used a blank sky ACIS
background events file (\texttt{acis7sD2000-01-29bkgrndN0003.fits})
when fitting the spectra, reprocessed to have the same gain file as
the dataset. We checked the count rate in the observation in a
8-10~keV band where there are likely to be very few source photons. It
matched the count rate in the blank sky background file to better than
0.5 per~cent.

\section{Analysis}
\begin{figure}
  \includegraphics[width=\columnwidth]{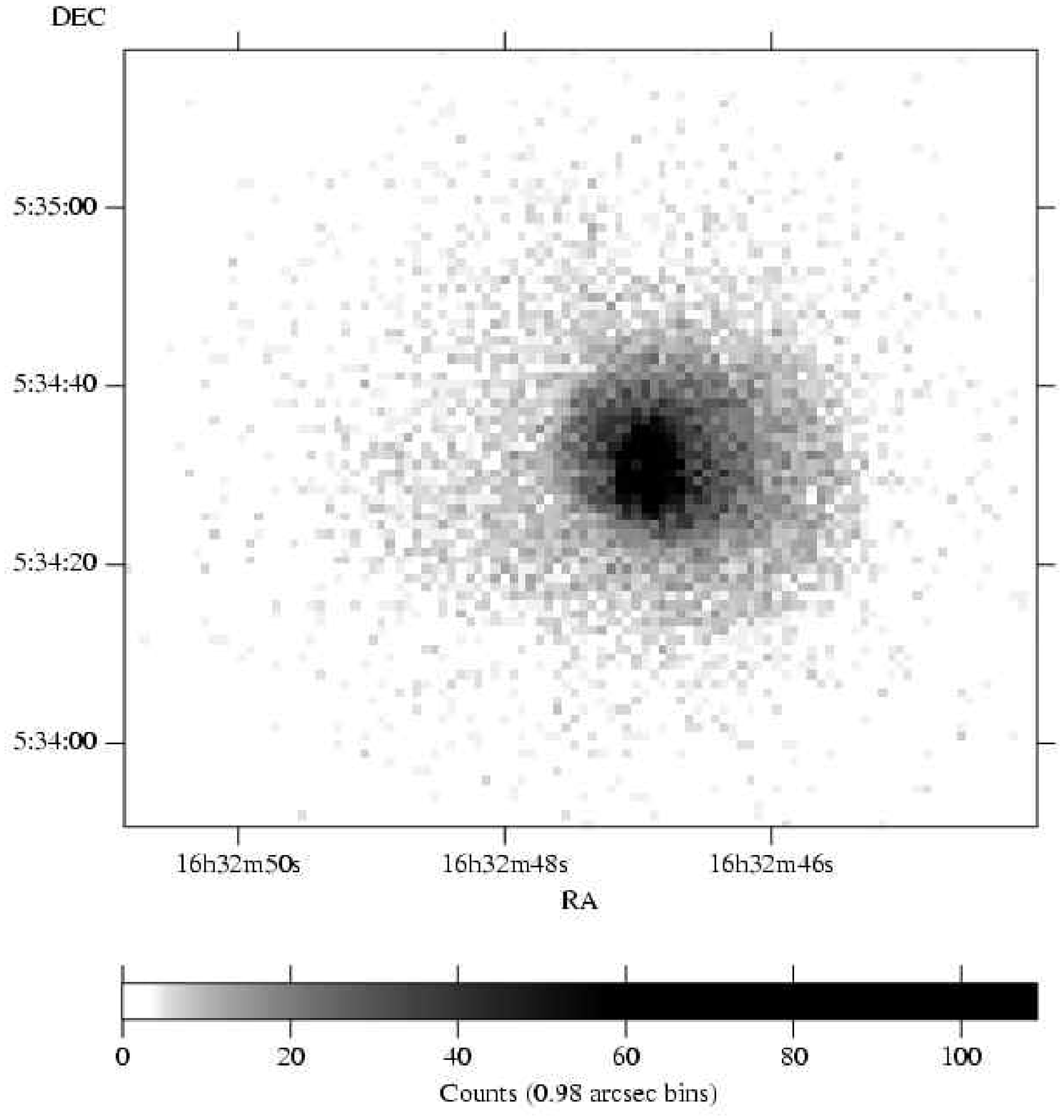}
  \includegraphics[width=\columnwidth]{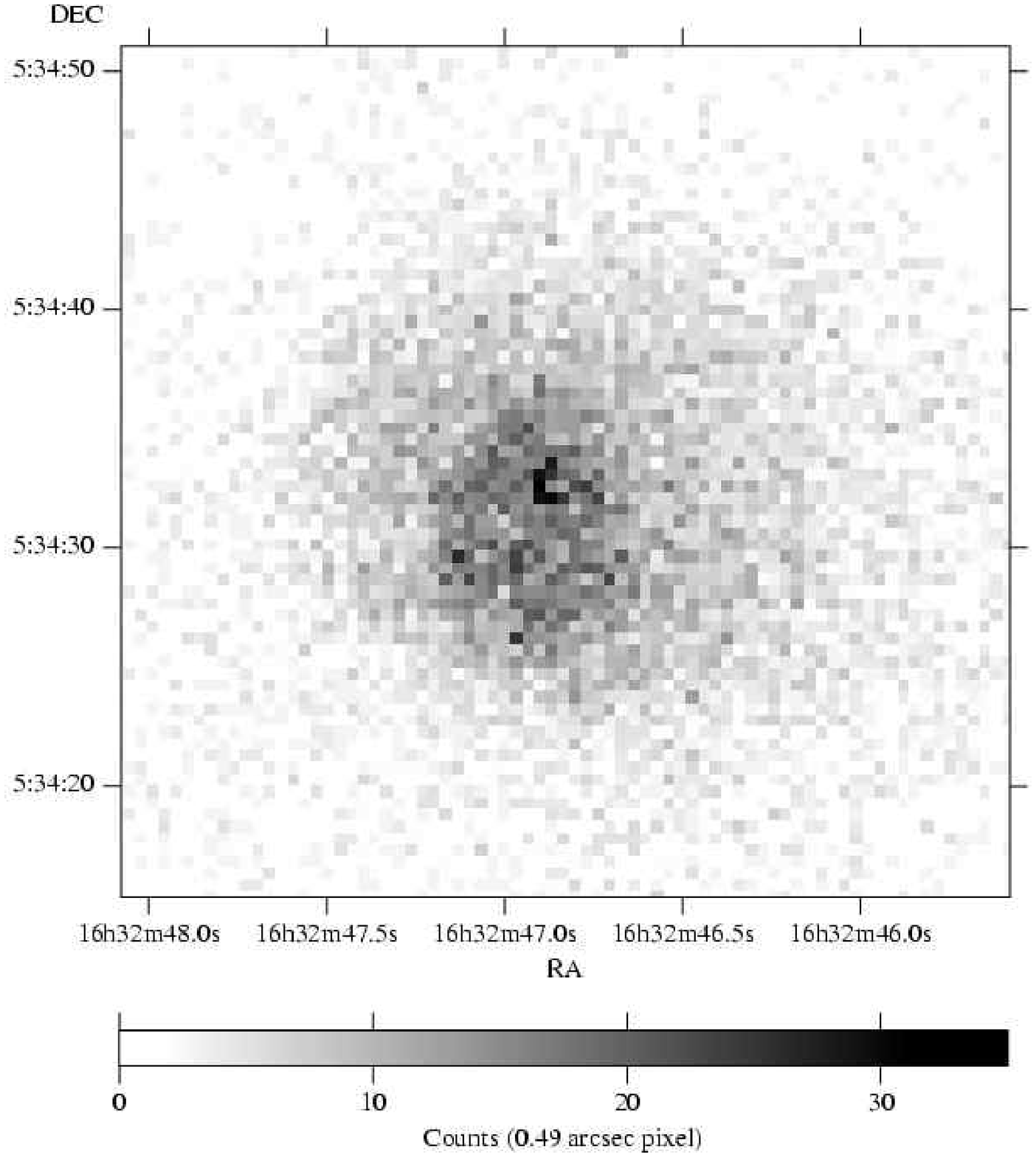}
  \caption{(Top) Image of the cluster between
    0.5 and 7 keV with 0.98 arcsec bins. (Bottom) Image of the central
    region with 0.49 arcsec bins.}
  \label{fig:images}
\end{figure}

Fig.~\ref{fig:images}~(top) shows an image of the central region of
the cluster using $\sim 1$~arcsec binning. In
Fig.~\ref{fig:images}~(bottom) we show an image with $\sim 0.5$~arcsec
binning of the innermost core. In these and the other images we
present, we have excluded visually identified point sources
(Table~\ref{tab:ptsrcs}).

\begin{table}
  \begin{tabular}{ll}
    RA & Dec \\ \hline
    16:32:56.8 & +05:34:59.9 \\
    16:32:51.3 & +05:34:53.5 \\
    16:32:53.1 & +05:32:02.3 \\
    16:33:00.2 & +05:37:32.9 \\
    16:32:58.0 & +05:37:38.8 \\
  \end{tabular}
  \caption{Excluded point sources (J2000).}
  \label{tab:ptsrcs}
\end{table}

Using a smoothed image we defined regions for spectral analysis using
a ``contour binning'' method (Sanders et al, in preparation).  The
algorithm defines regions with a signal to noise greater than a
threshold by ``growing bins'' in the direction on a smoothed map which
has a value closest to the mean value of those pixels already binned.
This technique defines bins which are matched to the surface
brightness profile of the object.  Additionally we applied constraints
on the regions to have edge lengths which are 3 times or less than
those of circles with the same area, preventing the bins from becoming
annular.

The smoothed map was created using an ``accumulative smoothing''
algorithm (Sanders et al, in preparation), smoothed to have a signal
to noise ratio of at least 8.  For each pixel, we find the signal to
noise ratio. If it is less than a set threshold, we add those pixels
of distance $\le$ 1 pixel from the starting point. We repeat,
increasing in radius, until the signal to noise threshold is reached.
The value of the smoothed pixel is the average of the pixels summed to
reach the signal to noise threshold.

We extracted spectra from the binned spatial regions for the dataset
and background dataset. In addition weighted responses (weighted by
the number of counts between 0.5 and 7~keV) and ancillary-responses
were created for each bin. We fitted the spectra in \textsc{xspec} by
minimising the C-statistic (Cash 1979).  The spectra were not grouped
and were fit between 0.6 and 8~keV. The model we fitted to each region
was a \textsc{mekal} emission spectrum (Mewe, Gronenschild \& van den
Oord 1985; Liedahl, Osterheld \& Goldstein 1995), absorbed by an
\textsc{phabs} absorption model (Balucinska-Church \& McCammon 1992).
In the fit the temperature, abundance, normalisation, and absorption
were free parameters.

\begin{figure*}
  \begin{tabular}{lr}
  \includegraphics[width=\columnwidth]{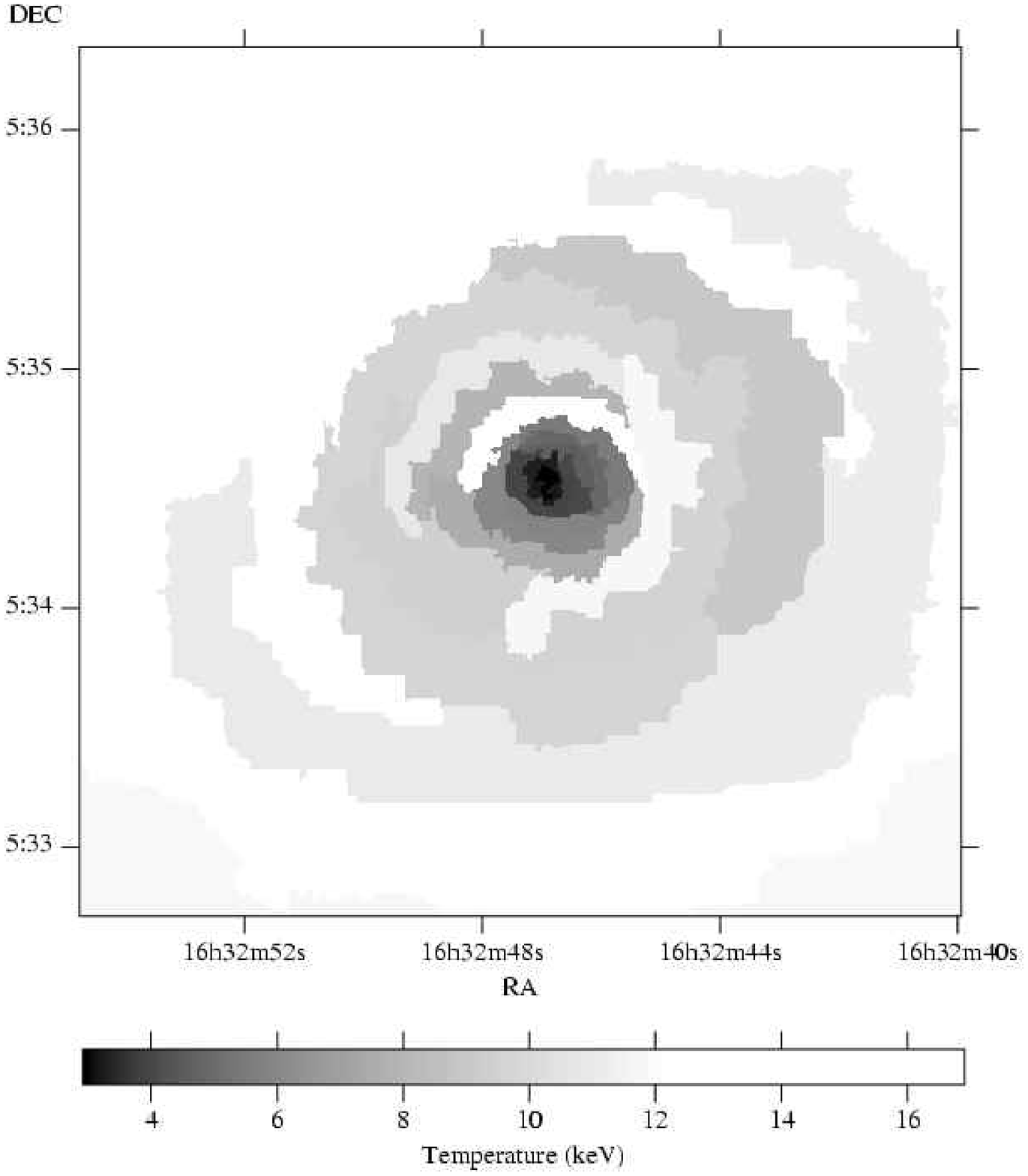}&
  \includegraphics[width=\columnwidth]{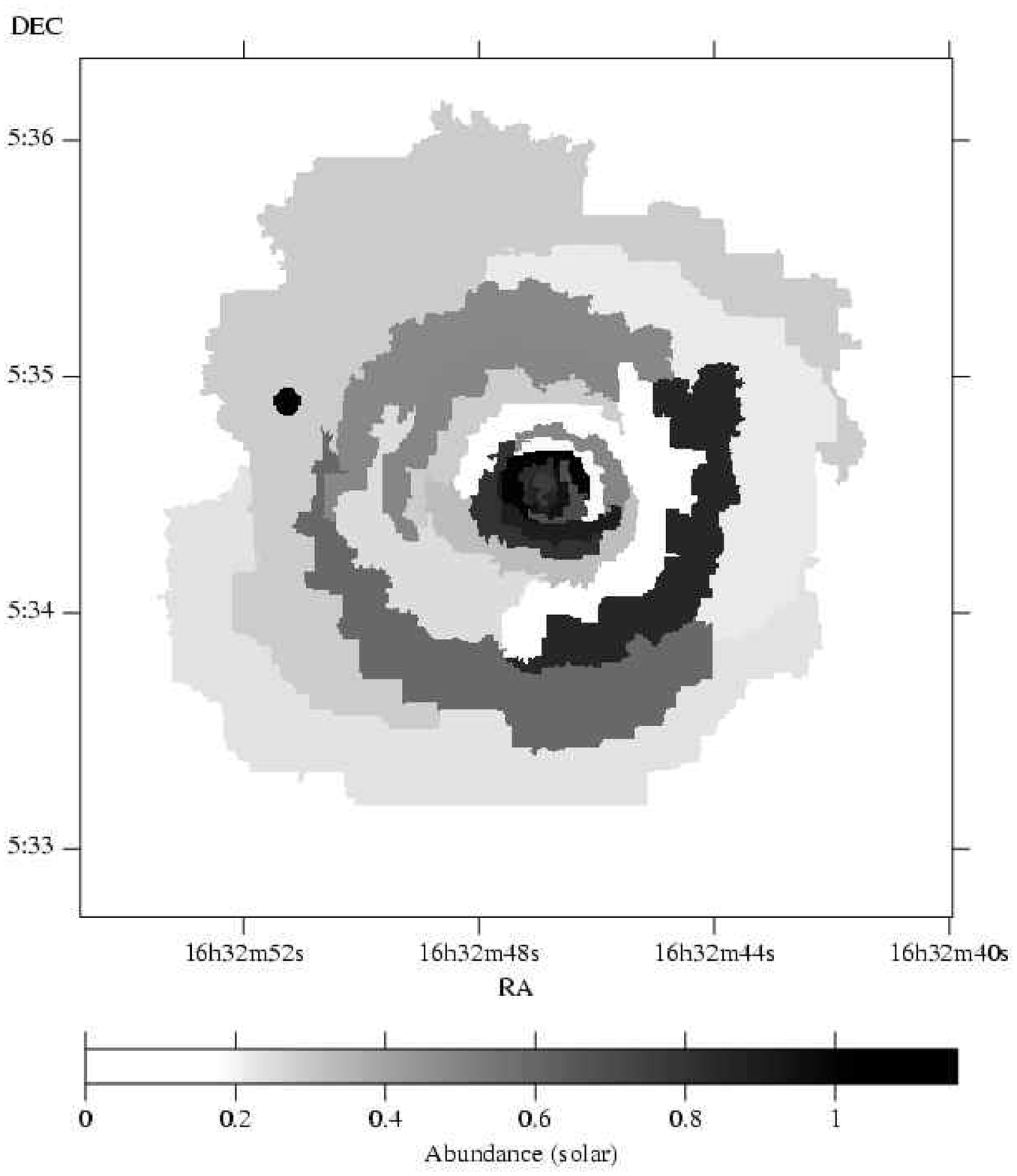}\\
  \includegraphics[angle=-90,width=\columnwidth]{fig2_bottomleft.eps}&
  \includegraphics[angle=-90,width=\columnwidth]{fig2_bottomright.eps}
  \end{tabular}
  \caption{Generated maps of the cluster using a signal to noise ratio
    $\ge 40$. (Upper Left) Temperature map of the cluster.
    (Bottom Left) Temperature profile of the regions in the above map.
    (Upper Right) Abundance map of the cluster. (Bottom Right)
    Abundance profile. The points are at the mean radius of each bin,
    whilst the horizontal error bars indicate the range of radii
    occupied. The vertical error bars show the 1-$\sigma$
    uncertainties.}
  \label{fig:maps}
\end{figure*}

The results for the spectral analysis were combined together to form a
map. In Fig.~\ref{fig:maps}~(left) we show a temperature map produced
by fitting spectra from regions with a signal to noise ratio of $\ge
40$ ($\gtrsim 1600$ counts per spectrum), and a profile of the
temperatures of each region, plotted by radius from the centre of the
cluster.  Similarly, in Fig.~\ref{fig:maps}~(right) we show an
abundance map and profile.

\begin{figure}
  \includegraphics[width=\columnwidth]{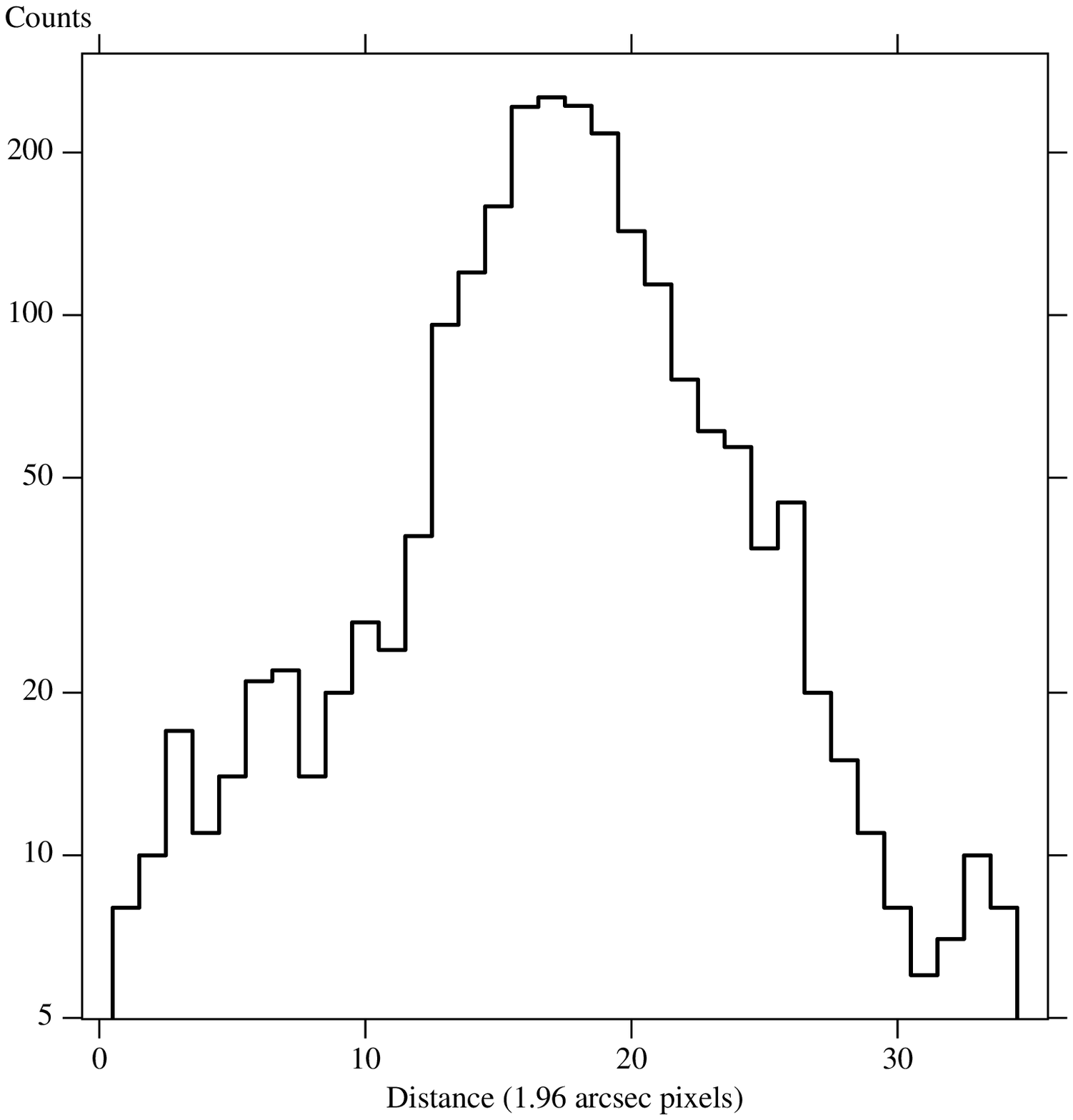}
  \caption{Profile over the core of the cluster from the NE to the
    SW, plotting the number of counts in 1.96 arcsec pixels. The cold
    fronts are at $\sim 12$ (NE) and 26 pixels (SW).}
  \label{fig:profile}
\end{figure}

\begin{figure}
  \includegraphics[width=\columnwidth]{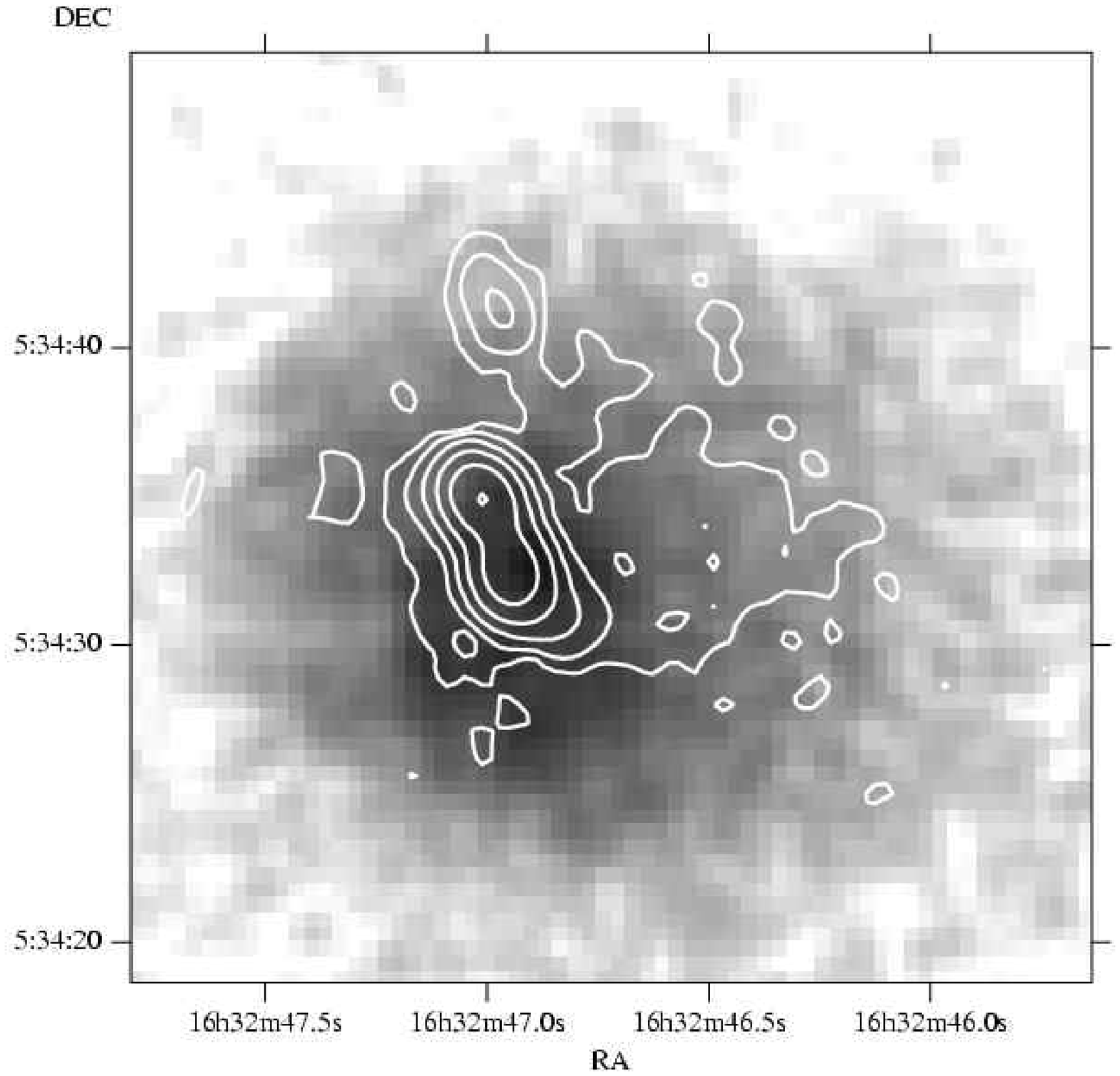}

  \caption{X-ray image between 0.5 and 7~keV, smoothed with a Gaussian
    of 0.49 arcsec, overlayed with contours from a VLA 1.4~GHz radio
    map of the central radio source. The 6 contours are spaced
    logarithmically between $10^{-4}$ and 0.022 Jy~beam${}^{-1}$. The
    beam size is $1.78 \times 1.39$ arcsec in position angle 33
    degrees.}
  \label{fig:radio}
\end{figure}

\begin{figure}
  \includegraphics[width=\columnwidth]{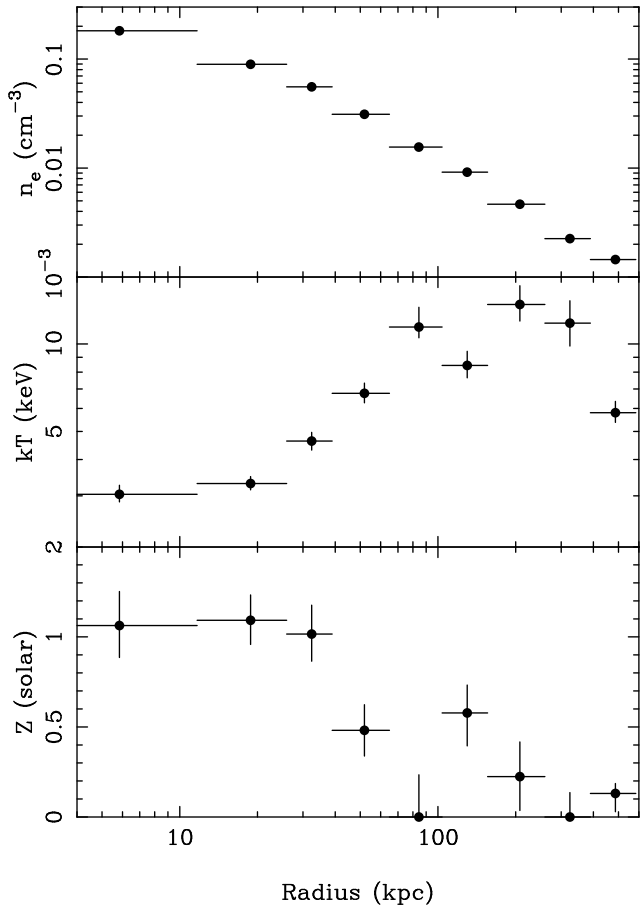}
  \caption{Electron density, temperature and abundance profiles in the
    cluster, accounting for projection. Error bars are 1$\sigma$.}
  \label{fig:profiles}
\end{figure}

The centre of the cluster contains an ellipsoidal core of dimensions
$7 \times 9$ arcsec ($19 \times 24$ kpc) in the east-west and
north-south directions. This core has a very flat surface brightness
profile (as seen in Fig.~\ref{fig:images}~[bottom] and
Fig.~\ref{fig:profile}).  There is a point source offset from the
centre of the core to the north-west by 2 arcsec, coincident in
position with the edge of the southern lobe of the radio source (See
Fig.~\ref{fig:radio}). The projected emission-weighted temperature of
the core region is $3.26 \pm 0.20$~keV with a high abundance of $1.11
\pm 0.20 \Zsun$ ($1\sigma$ uncertainties), found using a
\textsc{mekal} model.  From Fig.~\ref{fig:maps}~(right) there are
indications that there could be an abundance drop in the very centre.
A similar effect is seen in other clusters (e.g. Centaurus -- Sanders
\& Fabian 2002; Abell 2199 -- Johnstone et al 2002; Perseus - Sanders
et al 2004; Schmidt, Fabian \& Sanders 2002; Churazov et al 2003; NGC
4636 -- Jones et al 2002).  However the evidence is much less
significant when we later examine the profile accounting for
projection (Fig.~\ref{fig:profiles}).

Like most X-ray luminous clusters, Abell 2204 hosts a moderately
bright radio source.  We obtained short (122 and 30 min) VLA
observations of TXS 1630+056 at 1.4 and 5 GHz respectively on 1998
April 23 with the VLA in its A-configuration.  In Fig.~\ref{fig:radio}
we show the 1.4 GHz radio image in contours overlaid on the X-ray
image.  The radio source has an unusual morphology consisting of three
components, roughly aligned in the N-S direction over 10 arcsec, with
an extension to the west.  In addition larger structures are hinted at
in the NVSS image (Condon et al 1998). Such a disturbed radio
morphology is similar to that seen in the centres of other dense
clusters such as PKS 0745-191 (Taylor, Barton \& Ge 1994) and A2052
(Zhao et al. 1993).  It is also worth noting that in the 5 GHz VLA
observations the brightest component, with a peak flux density of 16
mJy, is $<$0.1\% polarised (3 $\sigma$).  Such a low polarised flux
density is also seen in the cluster PKS 0745-191 and is most likely
the result of very high Faraday rotation measure gradients produced by
a tangled cluster magnetic field (Taylor, Barton \& Ge 1994).  There
are no apparent holes in the X-ray emission corresponding to the main
northern radio lobe. This would be the case if the lobe is along the
line of sight but not in the X-ray bright ellipsoidal core.

\begin{figure}
  \includegraphics[width=\columnwidth]{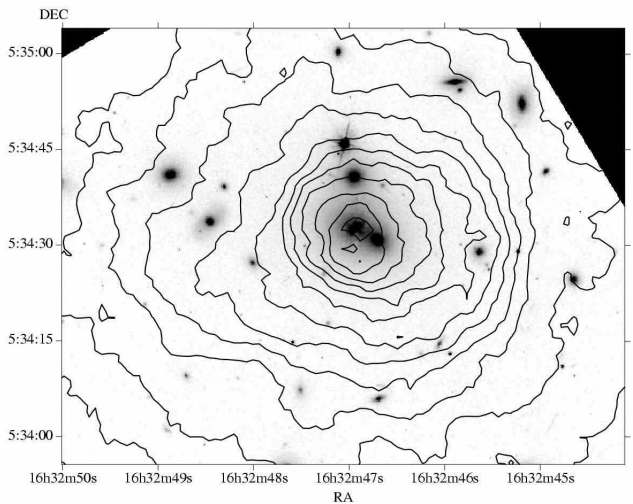}
  \caption{\emph{HST} image of the centre of the cluster with
    accumulatively smoothed X-ray contours overlaid. The \emph{HST}
    image is taken from the WFPC2 Associations catalogue,
    http://archive.stsci.edu/hst/wfpc2/ (association name U5A44101B)
    using the F606W filter. The X-ray image is an accumulatively
    smoothed image between 0.5 and 7~keV with a signal to noise ratio
    of 10, with 16 contours logarithmically spaced between 0.1 and 24
    counts per pixel.}
  \label{fig:hst}
\end{figure}

In Fig.~\ref{fig:hst} we show contours from an accumulatively-smoothed
X-ray image of the cluster on an optical image from the \emph{Hubble
  Space Telescope} (\emph{HST}). There are two closely separated
galaxies at the centre of the cluster separated by a distance of
4~arcsec. The northern-most galaxy is associated with the brightest
region of emission on the X-ray map in the core (although we find an
offset of $\sim1$~arcsec if the absolute astronometry is correct),
whilst the other galaxy to the south-west lies off the core.  The
northern-most galaxy has a fairly irregular morphology with a strong
dust lane extending from close to the core to the south, and other
possible regions of absorption lying around its core. It is unclear
whether the two galaxies lie in close proximity in the cluster or
whether they are only associated by line of sight. However the small
velocity difference between the galaxies ($249\kmps$; Jenner 1974)
suggests that they are closely associated.

The core is embedded in another region with a smooth surface
brightness profile and radius $\sim10$~arcsec. Within this central
plateau the core is displaced to the south. The central plateau has a
temperature of $3.7 \pm 0.2$ keV and an abundance of $0.85 \pm 0.15
\Zsun$.

\begin{figure*}
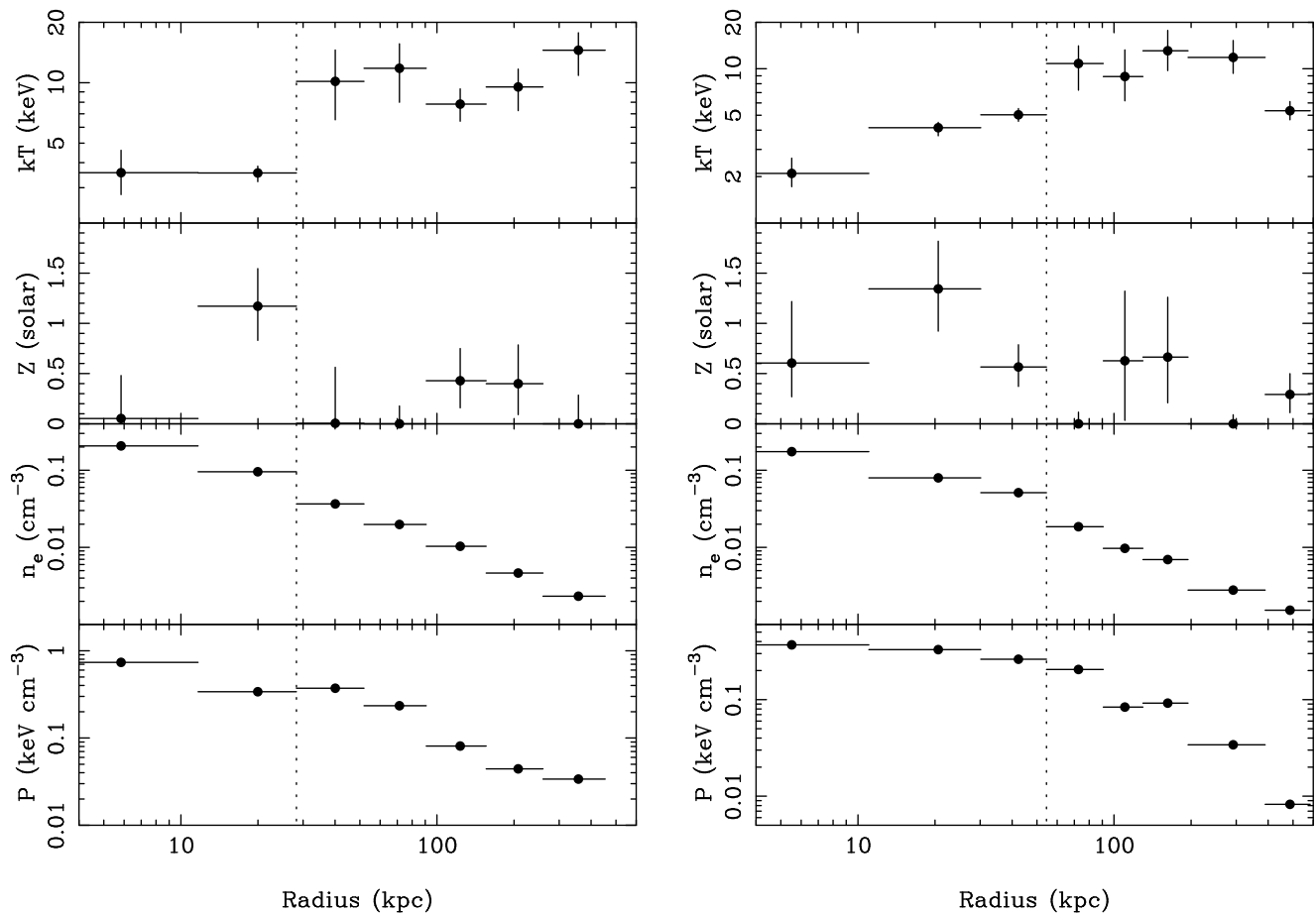

  \includegraphics[width=\columnwidth]{fig6_left.eps} \hspace{5mm}
  \includegraphics[width=\columnwidth]{fig6_right.eps}
  \caption{Temperature, abundance, density and pressure profiles
    across the two cold-fronts. The dotted line marks the radius of
    the cold front. (Left) The inner cold front to the north-east.
    (Right) The outer cold front to the west. Note that the pressure
    is continuous across both fronts.}
  \label{fig:cfprofile}
\end{figure*}

To the north-east of the central plateau there is an abrupt drop in
surface brightness by a factor of $\sim 2.5$. If we extract and fit
projected spectra either side of this jump in brightness, on the
innermost side the gas has a temperature of $4.1 \pm 0.6$ keV and
abundance of $0.97^{+0.35}_{-0.22} \Zsun$, but on the outside the
temperature is $9.3_{-2.0}^{+3.0}$~keV, with abundance $<0.6$
(1-$\sigma$). Therefore there is a jump in temperature by a factor of
$\sim 2.3$. We extracted spectra in sectors either side of the drop,
and fitted the \textsc{projct} model in \textsc{xspec} with an
absorbed single temperature \textsc{mekal} component. This model
accounts for projection by including the contribution of gas lying
outside each shell to the spectra. We allowed the temperature,
abundance and normalisation of each shell to be free, but tied the
absorption to be the same in each shell.
Fig.~\ref{fig:cfprofile}~(left) shows the temperature, abundance,
density and pressure profiles across the drop (whose radius is marked
with a dotted line). It can be seen that although the temperature
rises dramatically and the density drops across the drop, the pressure
remains constant. Therefore the drop is likely to be a cold front
(Markevitch et al 2000).

Extending from the west of the central plateau is a plume-like feature
which wraps clockwise around that plateau, decreasing in surface
brightness along its length, where the plume merges into the
surrounding diffuse emission. The plume is associated with a region of
lower temperature on the temperature map. To the west of the plume and
core is a significant drop in surface brightness. The temperature
beyond the drop is well fitted with a \textsc{mekal} model with a
temperature of $10.0_{-1.4}^{+1.8}$ keV, but inside is
$6.6_{-1.0}^{+1.8}$ keV. The surface brightness falls by a factor of
3.6, which is greater than at the previous cold front. We again
extracted and fitted spectra in sectors accounting for projection
(Fig.~\ref{fig:cfprofile}~[right]).  Similarly we found that the
pressure across the drop is continuous.  Therefore it is likely to be
another cold front.

The temperature map and profile (Fig.~\ref{fig:maps}~[left]) indicate
that the temperature rises with radius along an approximate power-law
until a radius of $\sim 100 \kpc$ when it starts to drop back down
again, before increasing again at around at $\sim 200\kpc$. This
behaviour is similar to that shown by the abundance map and profile
(Fig.~\ref{fig:maps}~[right]). The abundance rises from $\sim 0.8
\Zsun$ at the centre to $\sim 1.2\Zsun$, then declining steeply with
radius.  Where the temperature drops down slightly at $\sim 100\kpc$,
the abundance rises up, before falling away again.

In Fig.~\ref{fig:profiles} we show electron density, temperature and
abundance profiles made by fitting the \textsc{projct} model in
\textsc{xspec} with a single temperature \textsc{mekal} model,
accounting for projection assuming spherical symmetry. We tied the
absorption to be the same in each shell. The best fitting value of
$N_H$, $(3.8 \pm 0.2) \times 10^{20} \psqcm$, is close to the Galactic
value of $3.2\times10^{20} \psqcm$ (Dickey \& Lockman 1990). The
profiles generated from the temperature and abundance maps
(Fig.~\ref{fig:maps}) show very similar features to the deprojected
plot.  The centre of the cluster has a high electron density ($\sim
0.2 \pcmcu$), and flat abundance and temperature profile. We can also
see the ring of increased abundance and reduced temperature at a
radius of $\sim 100\kpc$. If we introduce a power-law component in the
innermost annulus, there is no significant improvement in the quality
of fit.

\begin{figure}
  \includegraphics[width=\columnwidth]{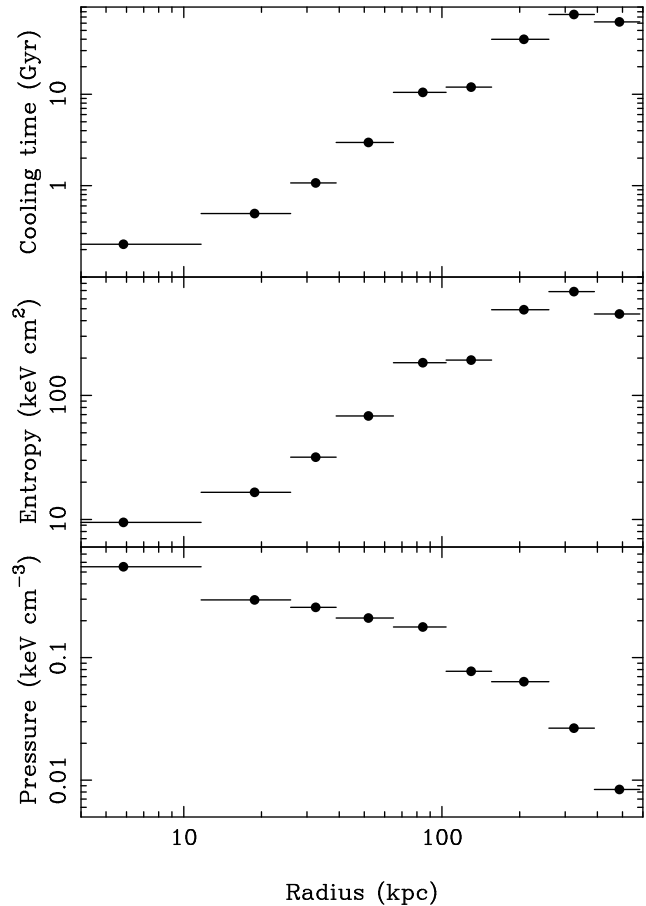}
  \caption{Mean radiative cooling time, entropy and electron 
    pressure profiles of the
    cluster, derived from Fig.~\ref{fig:profiles}.}
  \label{fig:tcool}
\end{figure}

From the above profiles the mean radiative cooling time
($t_\mathrm{cool} = \frac{5}{2} kT n / L$, where $L$ is the luminosity
per unit volume), entropy ($S=kT \: n_e^{-2/3}$) and electron pressure
($P=kT \: n_e$) profiles can be calculated (Fig.  \ref{fig:tcool}).
The plot shows that the cooling time of the gas within $\sim 10$~kpc
is short ($\sim 230$~Myr), increasing to the age of the universe at
around 100~kpc. If we include an extra \textsc{mkcflow} cooling flow
model component to the innermost deprojected region (cooling from the
temperature of the \textsc{mekal} component to zero, at the same
abundance), a 2-$\sigma$ upper limit for the cooling flow flux is $12
\Msunpyr$.  However this value depends on the absorption used. If we
allow the absorption to vary in each shell (note that this is not done
in a physically-consistent way, as the absorption is not applied in
projection), then there is a cooling flow limit of $97 \Msunpyr$
(2-$\sigma$) in the innermost shell, with a best fitting value of $40
\Msunpyr$. Instead of the cooling flow component we also tried an
additional powerlaw component. There was no significant reduction in
the fit-statistic when the component was added.

To further constrain the amount of cooling in this cluster, we
extracted a spectrum from the inner 50~kpc, and fit it with a
\textsc{mkcflow} model with the abundance, upper and lower temperature
free, but with the absorption fixed at the Galactic value similar to
that used on other clusters (Peterson et al 2003). The best fitting
model cooled from $8.0^{+0.7}_{-1.2}$~keV to $2.2^{+0.5}_{-0.2}$~keV
at a rate of $1090_{-130}^{+430} \Msunpyr$ with an abundance of $0.78
\pm 0.06 \Zsun$. If we introduced a further component cooling from the
lower temperature of the first component to zero, we found an upper
limit for the rate of cooling of $15\Msunpyr$ (2-$\sigma$).  It is
interesting to compare these values to Peres et al (1998) who measured
fluxes of $842^{+127}_{-82}$ and $843^{+245}_{-152} \Msunpyr$ with a
surface-brightness-deprojection technique using the PSPC and HRC on
\emph{ROSAT}, and Allen (2000) who spectrally measured
$2103^{+356}_{-378}\Msunpyr$ with \emph{ASCA}.

These results indicate that the temperature distribution of the gas
within the central 50~kpc is consistent with radiative cooling between
8 and 2~keV. Presumably some form of distributed heating is preventing
a full cooling flow from developing. The lack of an accumulation of
gas at a particular temperature (similar to the result found in the
Perseus cluster; Sanders et al 2004) means that the heating has to
operate over the full temperature range.  A residual central cooling
rate of at least several $\Msunpyr$, consistent with our spectra, is
needed to fuel the star formation rate observed (Crawford et al 1999).

Where the temperature decreases in the outer part of the cluster
($\sim 100\kpc$), it is interesting that the cooling time and entropy
only slightly change, but the pressure decreases by a factor of $2.3$.
Furthermore this radius is where the cooling time reaches the age of
the universe. At very large radii ($\gtrsim 400$~kpc) the temperature
of the cluster appears to drop dramatically to 6-7~keV. We do not
understand the cause of this apparent drop but it appears similar to
what is seen in Abell~3581 (Johnstone et al, submitted).

\section{Discussion}
There are several indications that the core of Abell~2204 is highly
disturbed. There is the morphology of its centre: the flat core, the
cold fronts, the central plateau, and the tail. In addition the
structure of the radio source is unusual.  Other evidence is provided
by the cool high abundance ring at around $100 \kpc$ and, possibly,
the binary appearance of the central galaxy.  Much of this can be
explained by a merger event. Despite this disruption the cooling time
of the centre is very short ($\sim 230$~Myr), similar to many other
centrally X-ray peaked clusters. Presumably this means that any merger
was not of equal mass but with a much smaller subcluster.

The presence of two cold fronts in Abell~2204 may indicate the
existence of substantial magnetic fields that suppress thermal
conduction and prevent Kelvin-Helmholtz instabilities from forming
(Vikhlinin, Markevitch \& Murray 2001).  Other indirect evidence for
cluster magnetic fields in Abell~2204 is the low fractional polarisation
seen at radio wavelengths. A merger has presumably left the central
galaxy/galaxies oscillating at the centre of the cluster potential,
thereby making the inner and outer cold fronts either side of the
central galaxy.

\section*{Acknowledgements}
ACF thanks the Royal Society for support. The authors are grateful to
Roderick Johnstone for discussions. The National Radio Astronomy
Observatory is operated by Associated Universities, Inc., under
cooperative agreement with the National Science Foundation.

\end{document}